\newcommand{\C}{$^\circ$C}
\newcommand{\mohm}{$\mu\Omega\,$cm}
\newcommand{\nohm}{n$\Omega\,$cm}

\documentclass{nature}

\usepackage[version=3]{mhchem} 
\usepackage[T1]{fontenc}       
\usepackage[version=3]{mhchem} 
\usepackage[T1]{fontenc}       
\usepackage{graphicx}          
\usepackage{dcolumn}           
\usepackage{bm}                
\usepackage{xcolor}            
\usepackage{amsmath}           
\usepackage{hyperref}          
\usepackage{tabularx}
\usepackage{lineno}

\title{
\begin{center}
{Evidence of a coupled electron-phonon liquid in NbGe$_2$}
\end{center}
}

\author{
Hung-Yu~Yang$^{1\,*}$,
Xiaohan~Yao$^1$,
Vincent~Plisson$^1$,
Shirin~Mozaffari$^2$,
Jan~P.~Scheifers$^3$,
Aikaterini Flessa Savvidou$^4$,
Gregory T. McCandless$^3$,
Mathieu~F.~Padlewski$^5$,
Carsten~Putzke$^5$,
Philip~J.~W.~Moll$^5$,
Julia~Y.~Chan$^3$,
Luis~Balicas$^{2,4}$,
Kenneth~S.~Burch$^1$,
Fazel~Tafti$^{1\,*}$
}

\begin{document}
\maketitle

\begin{affiliations}
 \item{Department of Physics, Boston College, Chestnut Hill, MA 02467, USA}
 \item{National High Magnetic Field Laboratory, Florida State University, Tallahassee, Florida 32310,USA}
 \item{Department of Chemistry and Biochemistry, University of Texas at Dallas, Richardson, TX 75080, USA} 
 \item{Department of Physics, Florida State University, Tallahassee, Florida 32306,USA}
 \item{\'{E}cole Polytechnique F\'{e}d\'{e}ral de Lausanne (EPFL), 1015 Lausanne, Switzerland, Laboratory for Atomic and Solid State Physics}
\end{affiliations}

\begin{abstract}
Whereas electron-phonon scattering typically relaxes the electron's momentum in metals, a perpetual exchange of momentum between phonons and electrons conserves total momentum and can lead to a coupled electron-phonon liquid with unique transport properties.
This theoretical idea was proposed decades ago and has been revisited recently, but the experimental signatures of an electron-phonon liquid have been rarely reported.
We present evidence of such a behavior in a transition metal ditetrelide, NbGe$_2$, from three different experiments.
First, quantum oscillations reveal an enhanced quasiparticle mass, which is unexpected in NbGe$_2$ due to weak electron-electron correlations, hence pointing at electron-phonon interactions.
Second, resistivity measurements exhibit a discrepancy between the experimental data and calculated curves within a standard Fermi liquid theory.
Third, Raman scattering shows anomalous temperature dependence of the phonon linewidths which fits an empirical model based on phonon-electron coupling.
We discuss structural factors, such as chiral symmetry, short metallic bonds, and a low-symmetry coordination environment as potential sources of coupled electron-phonon liquids.  
\end{abstract}

\pagebreak
\section{\label{sec:intro}Introduction}
The transport properties of metals with weak electron-electron (el-el) correlations are well described by the Fermi liquid theory and Boltzmann transport equation~\cite{abrikosov_fundamentals_2017}.
Within the standard Fermi liquid theory, the quasiparticle effective masses, Fermi velocities, and electron-phonon (el-ph) scattering rates can be computed reliably from first principles.
These quantities are then used to calculate the electrical, optical, and thermal properties of metals and semimetals with trivial and topological band structures~\cite{coulter_microscopic_2018,sundararaman_theoretical_2014,tong_comprehensive_2019} by using the Boltzmann transport equation and assuming momentum-relaxing collisions between electrons and phonons.
%
Historically, a deviation from this standard framework has been predicted if the momentum transfered from electrons to phonons through el-ph scattering would recirculate from phonons back to electrons (through so-called ph-el scattering), instead of being dissipated in the lattice through anharmonic ph-ph scattering (phonon decay)~\cite{gurzhi_hydrodynamic_1972,steinberg_viscosity_1958,wiser_electrical_1984}.

Recent theoretical works have shown the emergence of an electron-phonon liquid under such conditions, i.e. when ph-el scattering dominates the anharmonic ph-ph scattering~\cite{levchenko_transport_2020,huang_electron-phonon_2020}.
The electrical and thermal conductivities of a correlated electron-phonon liquid are predicted to be superior to conventional Fermi liquids, since the cooperative ph-el interactions are momentum-conserving~\cite{levchenko_transport_2020}. 
In addition, several distinct transport regimes with unconventional thermodynamic properties are predicted in electron-phonon liquids, but experimental progress is hindered by the lack of candidate materials~\cite{huang_electron-phonon_2020}.
%
Here we present mounting evidence of such a liquid in NbGe$_2$ from three distinct measurements, namely torque magnetometry, resistivity, and Raman scattering. 
We also discuss the structure-property relationships that lead to the observed behavior and propose design principles to create future candidate materials. 	

\begin{figure*}
 \centering
 \includegraphics[width=\textwidth]{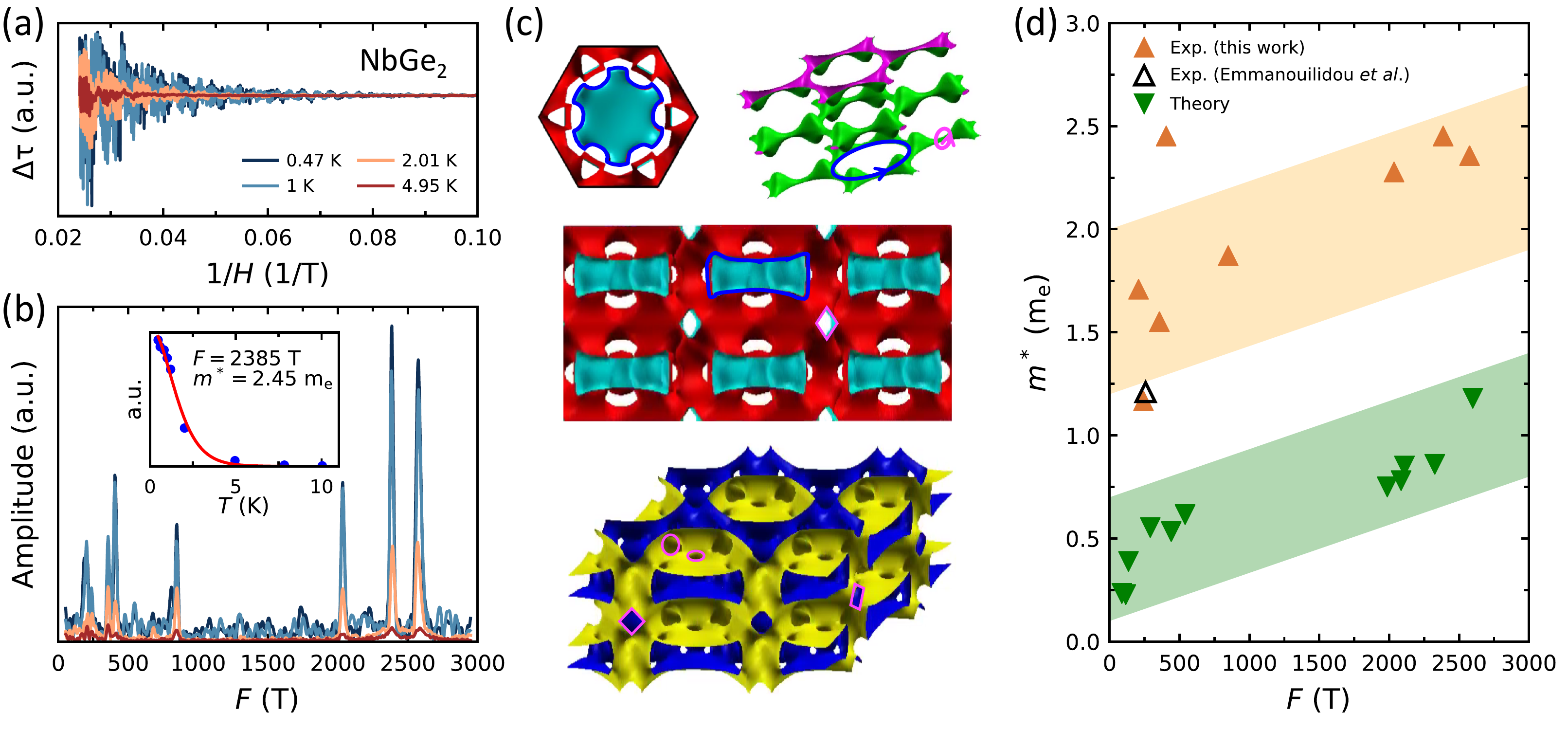}
 \caption{\label{fig:QO}
 \textbf{de Haas-van Alphen (dHvA) effect.}
 (\textbf{a}) de Haas-van Alphen (dHvA) oscillations as a function of inverse field plotted at four representative temperatures. Data were collected from a NbGe$_2$ crystal mounted on a piezoresistive cantilever.
 (\textbf{b}) The fast Fourier transform (FFT) of dHvA data. The inset shows a Lishitz-Kosevich fit to determine the effective mass at $F=2358$~T. 
 (\textbf{c}) Calculated Fermi surface of NbGe$_2$. The blue and magenta traces are cyclotron orbits as large as 2500~T and as small as 200~T, consistent with observed frequencies.
 (\textbf{d}) The experimental and theoretical quasiparticle masses plotted as a function of dHvA frequencies showing a three-fold enhancement in the experimental masses. The open data point is reproduced from Ref.~\cite{emmanouilidou_fermiology_2020}. The orange and green shaded areas highlight two standard deviations in the experimental and theoretical $m^*$ values, respectively.
 }
\end{figure*}

\section{\label{sec:qo}Quantum Oscillations}
In a Fermi liquid with weak el-el interactions, density functional theory (DFT) can be used to accurately compute the Fermi surface and effective mass of quasiparticles from first principles~\cite{muller_determination_2020}.
NbGe$_2$ seems to be just such a system: it is non-magnetic, does not have $f$-electrons, and is not close to a metal-insulator transition.
Therefore, it came as a surprise to find out the experimental values of the quasiparticle effective masses ($m^*$) were enhanced consistently beyond the DFT values across all branches of the Fermi surface.

We obtained the experimental $m^*$ values by measuring de Haas-van Alphen (dHvA) effect between 0.5 and 10~K, and from 0 to 41~T. 
The dHvA oscillations and their Fourier transform are plotted in Fig.~\ref{fig:QO}a and \ref{fig:QO}b, respectively.
The frequency ($F$) of each peak in Fig.~\ref{fig:QO}b is related to the extremal area ($A$) of a closed cyclotron orbit on the Fermi surface through the Onsager relation $F=\frac{\phi_0}{2\pi^2}A$.
For every orbit, the quasiparticle effective mass is evaluated by fitting the temperature dependence of the FFT peak intensity to a Lifshitz-Kosevich formula~\cite{shoenberg_magnetic_2009,yang_extreme_2017} (inset of Fig.~\ref{fig:QO}b and the Supplementary Fig.~S1 and Table~S1).

The Fermi surface of NbGe$_2$ in Fig.~\ref{fig:QO}c is calculated using density functional theory (DFT) and the theoretical $m^*$ values are obtained using a SKEAF program~\cite{rourke_numerical_2012}.
A comparison between the theoretical (DFT) and experimental (dHvA) $m^*$ values is presented in Fig.~\ref{fig:QO}d.
Both data sets increase uniformly with increasing frequency; however, the experimental values (orange) are three times larger than the theoretical ones (green) at all frequencies.  
As mentioned above, the el-el interactions must be weak in NbGe$_2$ since it is a non-magnetic metallic system without $f$-electron, and its Fermi surface comprises equal contributions from Ge-$p$/$s$ and Nb-$d$ orbitals (Fig.~S2).
Thus, the only viable explanation for such a systematic mass enhancement is a strong ph-el interaction.

\section{\label{sec:resistivity}Electrical Resistivity}
\begin{figure*}
 \centering
 \includegraphics[width=\textwidth]{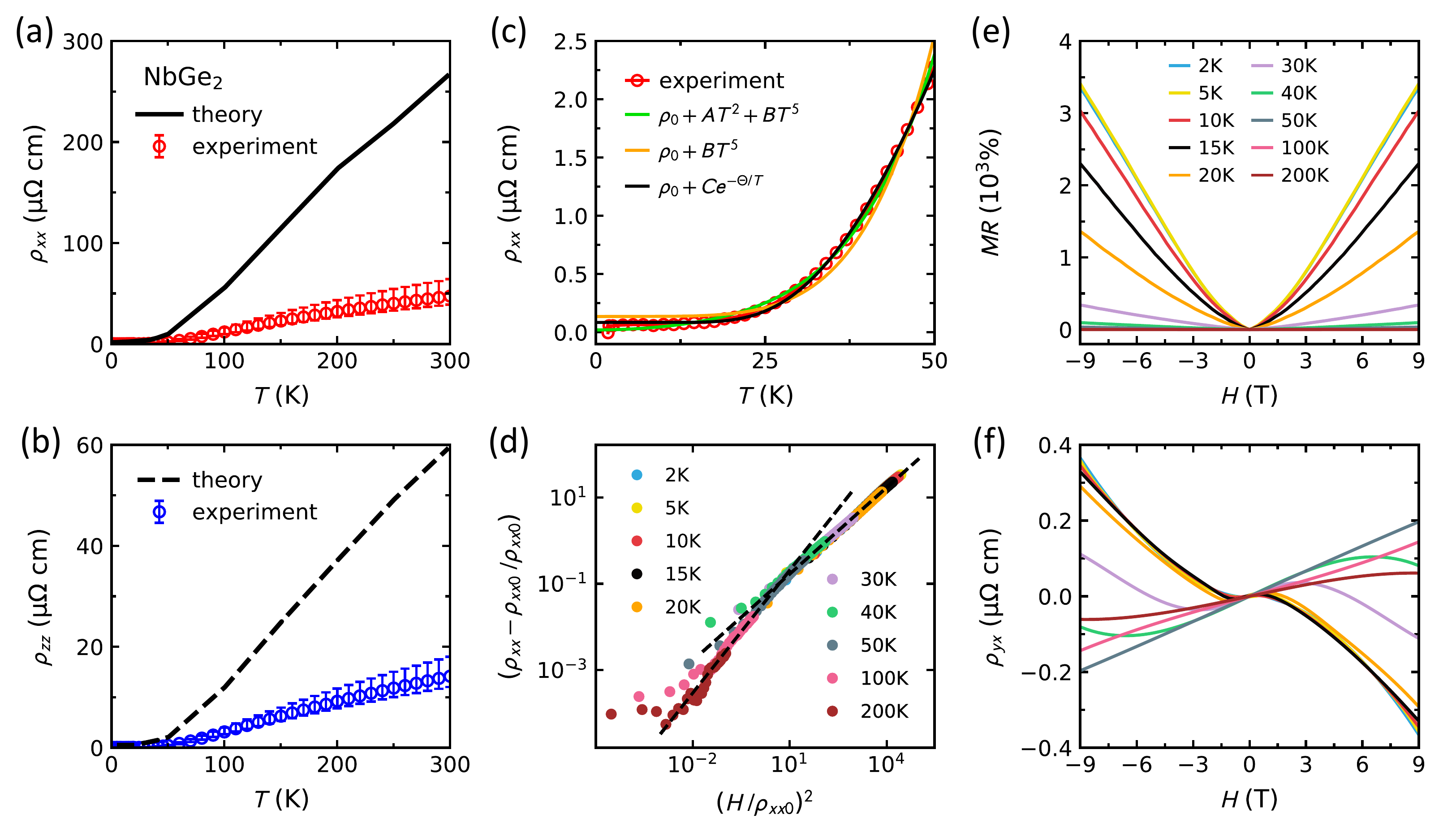}
 \caption{\label{fig:RT}
 \textbf{Electrical resistivity.}
 (\textbf{a}) The black lines show theoretical calculations of in-plane resistivity ($\mathit{\rho_{xx}}$) in NbGe$_2$ from Ref.~\cite{garcia_anisotropic_2020}, and the circles show the experimental data with error bars.
 (\textbf{b}) The same comparison is made for the out-of-plane resistivity ($\mathit{\rho_{zz}}$). Both $\mathit{\rho_{xx}}$ and $\mathit{\rho_{zz}}$ are measured on the same NbGe$_2$ sample with $RRR$=1030.
 (\textbf{c}) Different models are fitted to the $\mathit{\rho_{xx}}$ data below 50~K.
 (\textbf{d}) The Kohler scaling analysis shows a change of slope around 50~K.
 (\textbf{e}) Magnetoresistance MR~=~$[\rho(H)-\rho_0]/\rho_0$ as a function of field at several temperatures.
 (\textbf{f}) Field dependence of the Hall effect becomes nonlinear and changes sign below 50~K.
 }
\end{figure*}
The second evidence of a coupled electron-phonon liquid in NbGe$_2$ comes from resistivity measurements in Fig.~\ref{fig:RT}.
A recent theoretical work has calculated the resistivity curves of NbGe$_2$ by assuming a Fermi liquid ground state, evaluating momentum-relaxing el-ph lifetimes $\tau^{\mathrm{MR}}_{el-ph}(\mathbf{k})$ and electron velocities $v_\mathbf{k}$ for all bands, and plugging these values into the Boltzmann equation~\cite{garcia_anisotropic_2020}.
The results of this standard calculation are compared to the experimental curves in Fig.~\ref{fig:RT}a and \ref{fig:RT}b for in-plane ($\rho_{xx}$) and out-of-plane ($\rho_{zz}$) current directions, respectively.
The comparison reveals a six-fold overestimation of resistivity by theory in both directions.
To ensure the discrepancy is not due to uncertainties in sample geometry, we have also measured a standard mesoscopic device (Fig.~S3) fabricated by focused ion beam (FIB) with geometric uncertainties less than 5\%.
The six-fold discrepancy between theory and experiment can arise from either of el-el, el-defect, or ph-el interactions, but the first two can be ruled out here.
Not only el-el interactions are unlikely in NbGe$_2$ as mentioned earlier, but also they typically lead to a higher experimental resistivity than the theoretical curves, opposite to the observed behavior in Fig.~\ref{fig:RT}a,b.
Electron-defect scattering is irrelevant in NbGe$_2$ with a residual resistivity as small as $\rho_{0,xx}=55$~\nohm\ along $a$-axis and $\rho_{0,zz}=35$~\nohm\ along $c$-axis (the residual resistivity ratio $RRR>1000$).  
Thus, the only plausible source of this discrepancy is the ph-el interaction, which is theoretically predicted to enhance electrical conductivity beyond a standard Fermi liquid~\cite{levchenko_transport_2020}, consistent with our observations in Fig.~\ref{fig:RT}a,b.
It is important to note that the theoretical curves in Fig.~\ref{fig:RT}a,b (from Ref.~[\citenum{garcia_anisotropic_2020}]) have been calculated by including only the momentum-relaxing el-ph interactions wihtout considering the effect of ph-el and the recycling of momentum from phonons back to electrons.

To examine the el-el, el-ph and ph-el interactions further, we fit three different models to the low-temperature resistivity data ($\rho_{xx}$) in Fig.~\ref{fig:RT}c.
The black line is a fit to the phonon-drag model $\rho_{xx}=\rho_0+Ce^{-\Theta/T}$ that assumes dominant momentum-relaxing umklapp el-ph scatterings at high-$T$ and small-angle (quasi momentum-conserving) el-ph and ph-el scatterings at low-$T$~\cite{peierls_zur_1932,hicks_quantum_2012,protik_electron-phonon_2020}.
The fit yields $\Theta=155$~K, approximately one third of Debye temperature $\Theta_D=433$~K determined from the heat capacity measurements in Fig.~S4.
The orange line in Fig.~\ref{fig:RT}c is a fit to the Bloch–Gr\"{u}neisen model $\rho_{xx}=\rho_0+BT^5$, which does not fit the data properly, but it improves after adding a $T^2$ el-el scattering term and using $\rho_{xx}=\rho_0+AT^2+BT^5$ (green line).
Although the $T^2+T^5$ model fits the data, its coefficients, $A=2.98\times 10^{-4}$~\mohm~K$^{-2}$ and $B=5.19\times 10^{-9}$~\mohm~K$^{-5}$, do not make physical sense.
Using the $A$-coefficient of resistivity and the Sommerfeld coefficient from the heat capacity ($\gamma=6.2$~mJmol$^{-1}$K$^{-2}$ in Fig.~S4), we evaluate the Kadowaki-Woods ratio $R_{\textrm{KW}}=\frac{A}{\gamma^2}=7.7$~\mohm\,mol$^2$K$^2$J$^{-2}$ which is unreasonably large and comparable to the values in heavy fermions (about $10$~\mohm\,mol$^2$K$^2$J$^{-2}$)~\cite{jacko_unified_2009}.
This is inconsistent with the mild mass renormalization of a factor 3 in Fig.~\ref{fig:QO}d and the absence of $f$-electrons in NbGe$_2$.

Based on the above discussion, the activated $\rho(T)$ behavior in NbGe$_2$ is consistent with a transition from momentum-relaxing umklapp scattering to a momentum-conserving ph-el scattering regime below approximately 50~K.
Such a change of scattering length scale is confirmed by a Kohler scaling analysis on the field-dependence of resistivity in Fig.~\ref{fig:RT}d.
For this analysis, we use $\rho_{xx}(H)$ curves at several temperatures and plot $[\rho(H)-\rho_0]/\rho_0$ versus $(H/\rho_0)^2$.
The scaling plot reveals a change of slope at approximately 50~K (see the dashed lines in Fig.~\ref{fig:RT}d), which is consistent with a change of scattering length scale and emergence of an electron-phonon liquid below this temperatures. 
The magnetoresistance data (MR=100$\times (\rho(H)-\rho_0)/\rho_0$) used for the Kohler analysis are shown in Fig.~\ref{fig:RT}e.
Curiously, the Hall effect data in Fig.~\ref{fig:RT}f show a change of slope from positive to negative at approximately 50~K, where the Kohler analysis shows a change of slope.

\section{\label{sec:raman}Raman Scattering}
\begin{figure*}
 \centering
 \includegraphics[width=\textwidth]{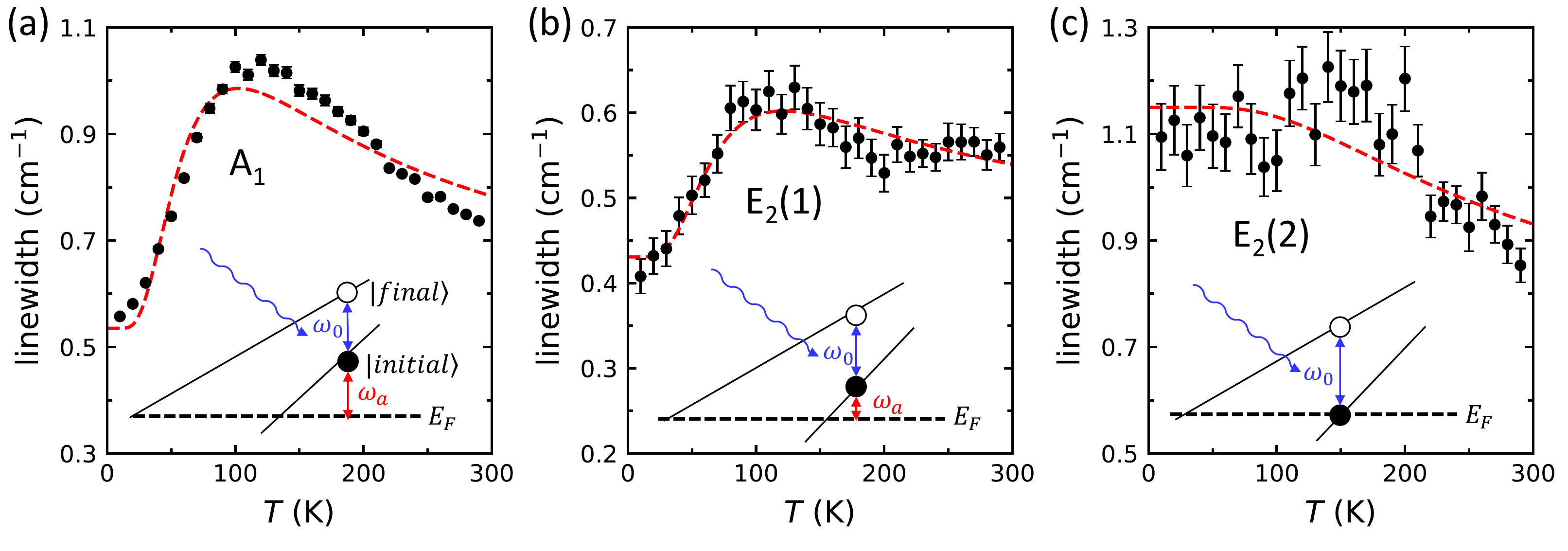}
 \caption{\label{fig:RAMAN}
 \textbf{Raman scattering.}
 Temperature dependence of Raman linewidth (proportional to inverse phonon lifetime) is plotted for three optical modes: (\textbf{a}) $A_1$, (\textbf{b}) $E_2 (1)$, and (\textbf{c}) $E_2(2)$. The red dashed line is a fit to Eq.~\ref{eq:gamma}. For each mode, the relationship between $\omega_0$ and $\omega_a$ (fit parameters in Eq.~\ref{eq:gamma}) is illustrated schematically.
 }
\end{figure*}
So far, we have focused on evidence of a correlated electron-phonon liquid in NbGe$_2$ by resorting to the electronic degrees of freedom (transport and dHvA data).
Now, we turn to the phononic degrees of freedom by examining the Raman linewidth as a function of temperature in Fig.~\ref{fig:RAMAN}.
Temperature dependent Raman scattering has recently been established as a sensitive tool for revealing the presence of dominant ph-el scattering~\cite{osterhoudt_evidence_2021}.
In NbGe$_2$, there are 16 modes with the mechanical representation $\Gamma_{\mathrm{opt.}}=A_1+2A_2+3B_1+2B_2+4E_1+4E_2$ that can be detected by Raman.
Typically, the finite phonon lifetime (hence finite linewidth) results from the anharmonic decay of optical to acoustic modes.
Because phonons are bosons, their linewidths are expected to scale with the Bose function $n_B(\omega,T)$ and increase with temperature - a behavior well captured by the Klemens model~\cite{klemens_anharmonic_1966,balkanski_anharmonic_1983}.
In stark contrast with the Klemens model, however, Fig.~\ref{fig:RAMAN} shows a non-monotonic temperature dependence in three representative modes that fits to a phenomenological model based on phonons decaying into electron-hole pairs. 
Specifically the linewidth is given by Fermi (instead of Bose) functions, according to 
\begin{equation}
\label{eq:gamma}
\Gamma(T)\propto n_F(\omega_a,T)-n_F(\omega_a+\omega_0,T)
\end{equation}
where $\omega_0$ is the phonon frequency, and $\omega_a$ is the energy difference between the electron's initial state and the Fermi energy in a phonon-mediated inter-band scattering~\cite{osterhoudt_evidence_2021}.
Note that Eq.~\ref{eq:gamma} is entirely phenomenological and independent of a specific theory.
Simply put, the $T$-dependences of optical phonons in NbGe$_2$ obey a Fermi (instead of Bose) function, which is possible only if the ph-el scattering dominates ph-ph scattering.

A physical picture of ph-el scattering emerges by comparing the fit parameters $\omega_0$ and $\omega_a$ in the insets of Fig.~\ref{fig:RAMAN}a,b,c.
For example, temperature dependence of the $A_1$ mode in Fig.~\ref{fig:RAMAN}a fits to Eq.~\ref{eq:gamma} with $\omega_a \approx \omega_0$, corresponding to a scenario where the initial electronic state is empty at $T=0$; it begins to populate with increasing temperature, and engages in ph-el scattering into an empty state (hole) via inter-band scattering. 
In other words, a phonon of frequency $\omega_0$ decays into an el-hole pair.
The initial increase of the phonon linewidth is due to increasing ph-el scattering rate with temperature.
At higher temperatures, however, the final state (hole) is also populated, so the phonons can no longer decay into an electron-hole pair and the linewidth decreases.
Thus, the initial increase and subsequent decrease of the linewidth is well-capture by Eq.~\ref{eq:gamma} in the entire temperature range.
A similar but less pronounced behavior is observed in Fig.~\ref{fig:RAMAN}b for the $E_2(1)$ mode, which fits to Eq.~\ref{eq:gamma} but with $\omega_a<\omega_0$.
Finally, the behavior in Fig.~\ref{fig:RAMAN}c for the $E_2(2)$ mode is described by Eq.~\ref{eq:gamma} with $\omega_a$=$0$, which means the electronic states are already populated at $T$=$0$ and the phonon linewidth only decreases with increasing temperature.

\section{\label{sec:results}Discussion}
\begin{figure}
 \centering
 \includegraphics[width=0.6\textwidth]{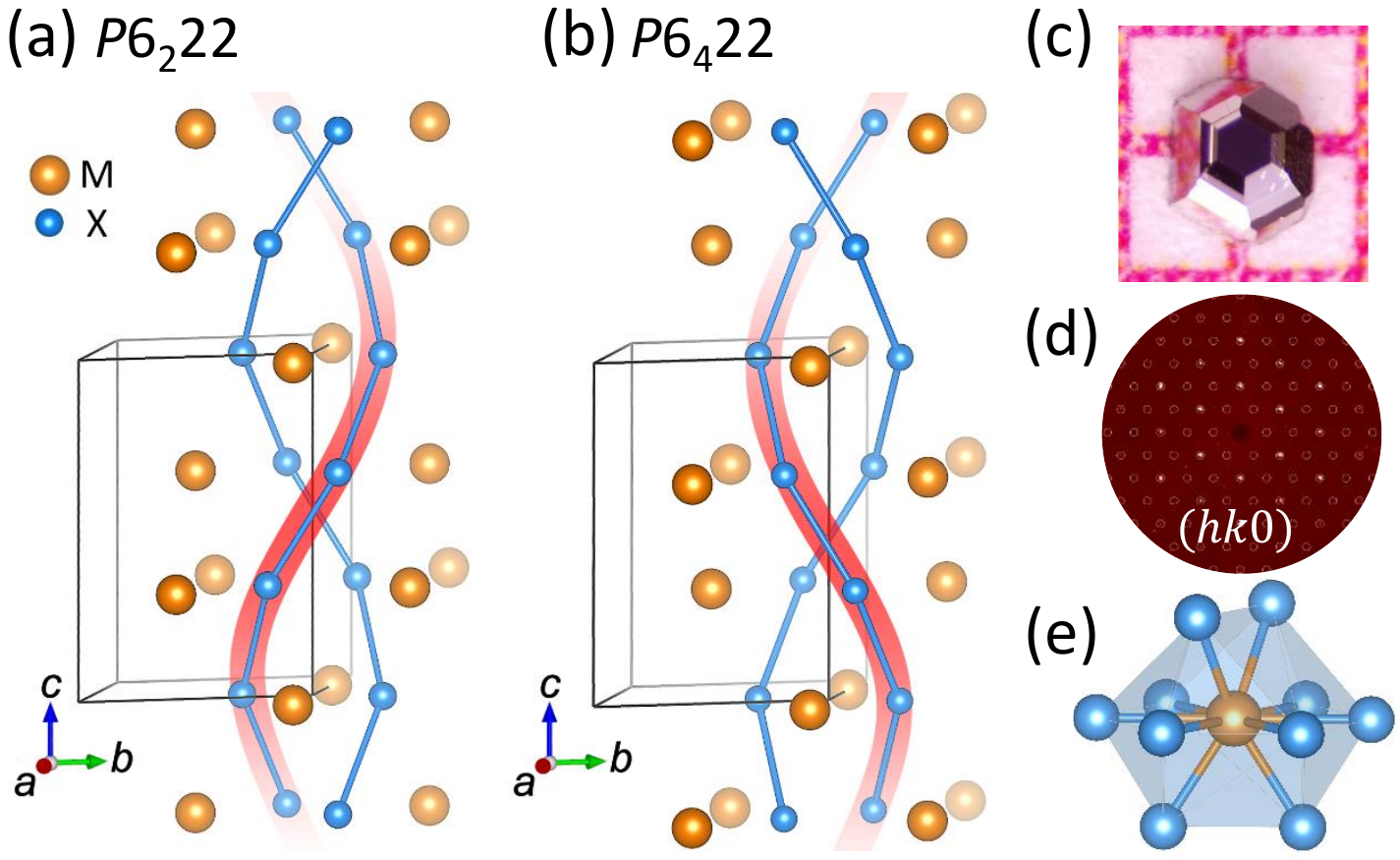}
 \caption{\label{fig:CIF}
 \textbf{Crystal structure of NbGe$_2$.}
 (\textbf{a}) The right-handed $6_2$ and (\textbf{b}) left-handed $6_4$ screw axes are compared in a generic MX$_2$ compound (e.g. NbGe$_2$) by showing the X-X bonds.
 (\textbf{c}) Picture of a millimeter size NbGe$_2$ crystal.
 (\textbf{d}) A precession image constructed from the single crystal X-ray diffraction data. The space group of NbGe$_2$ is $P6_222$ (right-handed).
 (\textbf{e}) The staggered dodecahedral coordination with 10 Ge atoms around each Nb atom in NbGe$_2$.
 }
\end{figure}

A few features in the structural chemistry of NbGe$_2$ may be responsible for the enhanced ph-el coupling in this material.
(i) NbGe$_2$ belongs to the C40 structural group, which is chiral due to the presence of a screw axis and the absence of an inversion center (Fig.~\ref{fig:CIF}a).
Two different chiralities (handedness) are observed among C40 structures~\cite{tanaka_refinement_2001}; the right-handed CrSi$_2$-type in space group $P6_222$ (\#180), and the left-handed NbSi$_2$-type in space group $P6_422$ (\#181).
The two structures can be distinguished by careful single crystal diffraction experiments (Fig.~\ref{fig:CIF}d and Table~S2).
Our crystallographic analysis in the Supplementary Information confirms the right-handed space group $P6_222$ in NbGe$_2$ crystals (Fig.~\ref{fig:CIF}a).
Specifically, a Flack parameter of 0 within the margin of error rules out enantiomeric twinning, which would correspond to the intergrowth of both chiralities (Table~S2)~\cite{parsons_use_2013}.  
Such a well-defined chirality is theoretically proven to stabilize Kramers-Weyl nodes in the electronic band structure~\cite{chang_topological_2018,tsirkin_composite_2017}, as confirmed in Fig.~S2 and elsewhere~\cite{garcia_anisotropic_2020,emmanouilidou_fermiology_2020}.
The Kramers-Weyl nodes may not be relevant to the electronic properties of NbGe$_2$ due to the large Fermi surface and carrier concentration of the order $10^{22}$ el/cm$^{3}$ (Fig.~S5 and Table~S3).
However, the lattice chirality may translate into chiral phonon modes which are known to affect transport properties in cuprate materials and control the el-ph coupling in WSe$_2$~\cite{grissonnanche_chiral_2020,zhu_observation_2018}. 
(ii) The short Nb-Ge and Ge-Ge bond lengths (2.7--2.9~\AA) in NbGe$_2$ maximize orbital overlaps and lead to extremely large residual resistivity ratios $RRR>1000$ and small residual resistivities $\rho_0<60$~\nohm\ (Fig.~2 and S6).
The large $RRR$ and small $\rho_0$ ensure that novel effects due to ph-el interactions are not masked by defect scattering.
The residual resistivity, carrier concentration ($10^{22}$ el/cm$^{3}$), and metallic bond lengths in NbGe$_2$ are comparable to those of PdCoO$_2$, which is also a hexagonal system with short Pd-Pd bond lengths of 2.8~\AA~\cite{shannon_chemistry_1971}.
PdCoO$_2$ is a candidate of electron hydrodynamics~\cite{moll_evidence_2016,mackenzie_properties_2017,nandi_unconventional_2018} possibly due to phonon-drag~\cite{hicks_quantum_2012,mackenzie_properties_2017}.
To understand whether NbGe$_2$ is close to an electron-phonon hydrodynamic regime~\cite{huang_electron-phonon_2020} will be an exciting future research direction.
(iii) The low-symmetry staggered dodecahedral coordination with 10 Ge around each Nb atom (Fig.~\ref{fig:CIF}e) creates nearly isotropic force constants, which in turn promote degenerate phonon states and a bunching between acoustic phonons.
It is shown theoretically that such an ``acoutic bunching effect'' limits the phase space for anharmonic decay of optical to acoustic phonons, leading to the dominance of ph-el over ph-ph scattering~\cite{lindsay_first-principles_2013,peng_phonon_2016}.
A similar effect is likely to suppress anharmonic phonon decays and produce high conductivity as observed in the Weyl semimetal WP$_2$, which also has a low-symmetry coordination environment~\cite{coulter_microscopic_2018,osterhoudt_evidence_2021}.
We propose the combination of a chiral lattice structure, short metallic bonds, and low-symmetry coordination complex as design principles to create new candidate materials for el-ph liquid~\cite{levchenko_transport_2020,huang_electron-phonon_2020}.

\pagebreak
\begin{methods}
 \subsection{Material Growth.}
Crystals of NbGe$_2$ were grown using a chemical vapor transport (CVT) technique with iodine as the transport agent.
The starting elements were mixed in stoichiometric ratios and sealed in silica tubes under vacuum with a small amount of iodine.
We found the best conditions to make high-quality samples was to place the hot end of the tube at 900\C\ under a temperature gradient of less than 10\C, and grow the crystals over a period of one month.
Polycrystalline samples were synthesized by heating as stoichiometric mixture of Nb and Ge powders at 900\C\ for three days.

\subsection{Transport and Heat capacity Measurements.}
The electrical resistivity was measured with a standard four-probe technique using a Quantum Design Physical Property Measurement System (PPMS) Dynacool.
The heat capacity was measured using the PPMS with a relaxation time method on a piece of polycrystalline sample cut from sintered pellets.
 
\subsection{X-ray Diffraction.}
Single crystal X-ray diffraction data was obtained at room temperature using a Bruker D8 Quest Kappa single crystal X-ray diffractometer operating at 50~kV and 1~mA equipped with an I$\mu$S microfocus source (Mo-K$_\alpha$, $\lambda$=0.71073~\AA), a HELIOS optics monochromator and PHOTON II detector. 
%
The structure was solved with the intrinsic phasing methods in SHELXT~\cite{sheldrick_crystal_2015}. 
%
No additional symmetries were found by the ADDSYM routine and the atomic coordinates were standardized using the STRUCTURE TIDY routine~\cite{gelato_structure_1987} of the PLATON~\cite{spek_single-crystal_2003} software as implemented in WinGX 2014.1~\cite{farrugia_wingx_1999}.

\subsection{Raman Scattering.}
Raman spectra were collected in a backscattering mode using a 532~nm Nd:YAG laser with incident power 200$\mu$W focused to a spot size of 2~$\mu$m in a Montana Instruments cryo-station~\cite{tian_low_2016}.
Polarization dependence for symmetry identification was performed via the rotation of a Fresnel rhomb which acts as a half-waveplate. 
The fitting of the phonon features to extract linewidths was performed using a Levenburg-Marquardt least squares fitting algorithm. 
Phonons were fit using a Voigt profile, wherein a Lorentzian representing the intrinsic phonon response is convoluted with a Gaussian to account for any broadening induced by the system

\subsection{de Haas-van Alphen (dHvA) Experiment.}
The magneto-quantum oscillation experiments under continuous fields up to 41~T were performed at the National High Magnetic Field Laboratory in Tallahassee, Florida. 
Temperature and angular dependences of the oscillations were examined to reveal the effective mass and dimensionalities of the Fermi surfaces of the samples. 
The de Haas–van Alphen effect in the magnetic torque were measured using piezoresistive cantilever technique (Piezo-resistive self-sensing $300\times100~\mu$m cantilever probe, SCL-Sensor.Tech.). 
A $^3$He cryostat in combination with a rotating probe was used for high-field experiments at temperatures down to 0.35~K.

\subsection{Density Functional Theory (DFT) Calculations.}
DFT calculations using the linearized augmented plane-wave (LAPW) method were implemented in the WIEN2k code \cite{blaha_wien2k_2018} with the Perdew-Burke-Ernzerhof (PBE) exchange-correlation potential \cite{perdew_generalized_1996} plus spin-orbit coupling (SOC).  
The basis-size control parameter was set to $\text{RK}_{\text{max}} = 8.5$ and $20000$ k-points were used to sample the k-space. 
Using DFT calculations as input, the Supercell K-space Extremal Area Finder (SKEAF) program~\cite{rourke_numerical_2012} was applied to find dHvA frequencies and effective masses of different Fermi  pockets.
\end{methods}



\pagebreak
\section*{References}
\bibliographystyle{naturemag} 
\bibliography{Yang_18feb2021}


\pagebreak
\begin{addendum}
 \item[Acknowledgments]
 F.T. thanks D.~Broido, A.~Levchenko, and A.~Lucas for helpful discussions.
 F.T. and H.-Y.Y. acknowledge funding by the National Science Foundation under Award No. NSF/DMR-1708929.
 Work done by V.P. was supported by the US Department of Energy (DOE), Office of Science, Office of Basic Energy Sciences under award no. DE-SC0018675. 
 K.S.B is grateful for the support of the Office of Naval Research under Award number N00014-20-1-2308. 
 J.Y.C acknowledges funding by the National Science Foundation under Award No. NSF/DMR-1700030.
 L.B. is supported by the US-DOE, BES program through award DE-SC0002613. The National High Magnetic Field Laboratory is supported by the National Science Foundation through NSF/DMR-1644779 and the State of Florida.
 C.P. and P.J.W.M. were supported by the European Research Council (ERC) under the European Union’s Horizon 2020 research and innovation programme (grant agreement No 715730).
 \item[Author Contributions] H.-Y.Y. and X.Y. grew the crystals, carried transport and heat capacity measurements, and performed DFT calculations. L.B. and S.M. performed dHvA experiments. J.P.S., G.T.M, and J.Y.C. analyzed the single crystal X-ray diffraction and crystal orientation. M. F. P., C.P., and P.J.W.M fabricated the FIB device. V.P. and K.S.B. performed Raman scattering. H.-Y.Y and F.T. wrote the manuscript.
 \item[Competing interests] The authors declare no competing interests.
 \item[Supplementary information] is available online.
 \item[Correspondence and requests for materials] should be addressed to H.-Y.Y. \&\ F.T.~(email: yanghw@bc.edu \&\ fazel.tafti@bc.edu).
\end{addendum}

\end{document}